\RequirePackage{ifpdf}
\documentclass[letterpaper]{article}
\usepackage{jheppub}
\usepackage{xcolor}
\usepackage{amsmath}
\usepackage{epsfig}
\usepackage{caption}
\usepackage{subcaption}
\usepackage{soul}
\pdfoutput=1

\usepackage{colortbl}
\usepackage{color}
\usepackage{tcolorbox}

\newcommand{\roughly}[1]{\mathrel{\raise.3ex\hbox{$#1$\kern-0.85em
\lower1ex\hbox{$\sim$}}}}

\newcommand{\gsim}{\roughly>}

\newcommand{\be}{\begin{equation}}
\newcommand{\bee}{\begin{equation}}
\newcommand{\ee}{\end{equation}}
\newcommand{\beea}{\begin{eqnarray}}
\newcommand{\eea}{\end{eqnarray}}
\newcommand{\bea}{\begin{eqnarray}}

\def\nott#1{\setbox0=\hbox{$#1$}                
   \dimen0=\wd0                                 
   \setbox1=\hbox{/} \dimen1=\wd1               
   \ifdim\dimen0>\dimen1                        
      \rlap{\hbox to \dimen0{\hfil/\hfil}}      
      #1                                        
   \else                                        
      \rlap{\hbox to \dimen1{\hfil$#1$\hfil}}   
      /                                         
   \fi}                                         %

\def\uxsl{\hbox{/\kern-.4000em$u$}}
\def\uxslsm{\hbox{\smaller/\kern-.5600em$u$}}
\def\pxpsl{\hbox{/\kern-.5000em$p$}}
\def\epssl{\hbox{/\kern-.5600em$\epsilon$}}
\def\delsl{\hbox{/\kern-.7000em$\nabla$}}
\def\lxpsl{\hbox{/\kern-.5600em$l$}}
\def\kxpsl{\hbox{/\kern-.5600em$k$}}
\def\qxpsl{\hbox{/\kern-.3900em$q$}}

\def\pref#1{(\ref{#1})}
\def\exd{{\rm d}}

\def\cO{{\cal O}}

\def\cR{{\cal R}}

\def\cV{{\cal V}}

\def\ssD{{\scriptscriptstyle D}}

\def\ssN{{\scriptscriptstyle N}}

\def\UV{{\scriptscriptstyle UV}}

\def\KK{{\scriptscriptstyle KK}}

\def\Li{{\rm Li}}

\newcommand{\bZ}{{\bf Z}}

\title{Perils of Towers in the Swamp:\\ Dark Dimensions and the Robustness of EFTs}

\author{C.P.~Burgess${}^{1,2,3}$ and F.~Quevedo${}^{2,4}$ \\ 
{\it 
${}^1$ Department of Physics \& Astronomy, McMaster University\\ 
\qquad 1280 Main Street West, Hamilton ON, Canada.\\
${}^2$ Perimeter Institute for Theoretical Physics\\
\qquad 31 Caroline Street North, Waterloo ON, Canada.\\
${}^3$ School of Theoretical Physics, Dublin Institute for
  Advanced Studies,\\ 
\qquad 10 Burlington Rd., Dublin, Co. Dublin, Ireland\\
${}^4$ DAMTP, University of Cambridge, Wilberforce Road,  Cambridge, CB3 0WA, UK.
}
}

\date{\today}

\abstract{Recently there has been an interesting revival of the idea to use large extra dimensions to address the dark energy problem, exploiting the (true) observation that towers of states with masses split, by $M^2_\ssN =  f(N) m^2,$  with $f$ an unbounded function of the integer $N$, sometimes contribute to the vacuum energy only an amount of order $m^\ssD$ in $D$ dimensions. It has been argued that this fact is a consequence of swampland conjectures and may require a departure from Effective Field Theory (EFT) reasoning. We test this claim with calculations for Casimir energies in extra dimensions. We show why the domain of validity for EFTs ensures that the tower spacing scale $m$ is {\it always} an upper bound on the UV scale for the lower-energy effective theory; use of an EFT with a cutoff part way up a tower is not a controlled approximation. We highlight the role played by the sometimes-suppressed contributions from towers in extra-dimensional approaches to the cosmological constant problem, old and new, and point out difficulties encountered in exploiting it. We compare recent swampland realizations of these arguments with earlier approaches using standard EFT examples, discussing successes and limitations of both.}


\begin{document}
\maketitle

\section{Towers of states and vacuum energy}
\label{sec:Towers}

A pressing theoretical question of our day is the zero-body problem: what is the energy density of the vacuum and why does this energy gravitate so little compared with theoretical expectations \cite{Zeldovich:1967gd, Weinberg:1988cp, Polchinski:2006gy, Burgess:2013ara}? Much has been thought and written about this problem, and a recent approach \cite{Montero:2022prj} builds on the following observation (based on arguments from swampland\footnote{The core assertion within the swampland program is that there exist otherwise sensible EFTs that do not have UV completions. To the extent that this paper bears on swampland issues, its spirit is not to use these conjectures as inputs, but to explore calculable implications of extra-dimensional physics (from which the conjectures have partially been abstracted) to help identify those that agree/disagree with the conjectures. This seems useful for assessing the evidence for/against the core assertion. } conjectures): 
\begin{quotation}
\noindent
In $D$ spacetime dimensions an infinite tower of states that are spaced by an energy scale of order $m$ -- for instance $M_\ssN^2 = f(N) m^2$ with $f(N) = N$ or $N^2$ or $\sqrt{N(N+1)}$ and so on, for integer $N$ (say) -- can naturally contribute a vacuum energy that is of order $\rho_\ssD \sim m^\ssD$. 
\end{quotation}
\noindent
This is at first sight a remarkable assertion because normally each element of the tower would be expected to contribute by an amount $\delta \rho_\ssD \sim M_\ssN^\ssD$ and so the sum over $N$ would seem to lead to a divergent quantity
\be 
   \rho_\ssD =  m^\ssD \sum_{N=0}^\infty c_\ssN \left[ \frac{f(N)}{4\pi} \right]^{D/2} \qquad \hbox{for} \quad c_\ssN \sim \cO(1) \,.
\ee   
The above assertion means the divergent sum counter-intuitively turns out to be order unity. At face value this appears to be a dramatic suppression relative to naive EFT reasoning.

There is nevertheless good evidence that the above claim is true, partly because it is not in itself a new observation. The novelty is its use to argue for the possible breakdown of EFT methods, and whether such towers can help identify low-energy situations that depend unusually strongly on the nature of gravity's UV completion, with the above assertion argued to provide evidence for a breakdown in EFT reasoning. If true this would strengthen the motivation for various (swampland) conjectures that grope towards an alternative framework for understanding the low-energy world without EFTs. 

Before discussing more broadly the role played by the above assertion in approaches to the cosmological constant problem, we first briefly describe the evidence for its validity and why it does not indicate a breakdown of EFT reasoning. 

\subsection{Vacuum energies from towers}

One line of evidence comes from string theory -- as emphasised in \cite{Montero:2022prj} -- for which $m \sim M_s$ is the string scale and the tower in question consists of excited states of a relativistic string (or superstring). One-loop vacuum energies have been known since the 1980s to be ultraviolet finite and in $D$ dimensions to be of order $\rho_\ssD \sim M_s^{\ssD}$ \cite{Polchinski:1985zf, Dienes:1995pm} (see also \cite{Alvarez-Gaume:1986ghj, Dixon:1986iz} for concrete non-supersymmetric string theories). In particular, although the full string result can be written in a way that looks like a naive mode-by-mode sum of vacuum energies for each string level once these are written using the heat-kernel formalism (see for instance \cite{HeatKernel}), the total result actually differs from this naive result because invariance under modular transformations of the string world sheet nontrivially restricts the heat-kernel integration regime in a way that excludes the dangerous UV-sensitive contributions.  

It is tempting to think from the string example that the UV finiteness of string theory plays an important role in this argument, but this is not true for the second line of evidence coming from dimensional reduction. In this class of examples the tower of interest consists of Kaluza-Klein modes for extra-dimensional fields, such as those arising when gravity or supergravity is dimensionally reduced from higher to lower dimensions. In this case the role of the tower spacing $m$ is played by the Kaluza-Klein scale $M_\KK$, which in simple examples scales like the inverse of extra-dimensional `size' $L$: $M_\KK \sim 1/L$. Vacuum energies can be explicitly computed for compactifications down to four dimensions using the higher dimensional theory, giving $\rho_4 \sim M_\KK^4$ (as opposed to being proportional to $M^4$ where $M$ is the UV scale in the higher dimensions, like the string scale or the scale of heavy higher-dimensional particle masses). 

For example, for six-dimensional theories explicit calculations of the Casimir energy for (untwisted) 6D massive scalar field compactified on a 2-torus \cite{Ponton:2001hq, Ghilencea:2005vm}  give a 4D vacuum energy\footnote{Extensions to  orbifolds were also computed in \cite{Ponton:2001hq, Ghilencea:2005vm} with similar properties.}
\be \label{6to4masslessC}
   \rho_4  = -\frac{1}{\cV^2} \, \bigg\{
    \frac{4\pi^3 U_2^3}{945}
    + \frac{3 \,\zeta(5)}{2\pi^2 U_2^2}
    +2  \sum_{k =1}^\infty \Big[ k^2\, \Li_3(q^{k})
    \phantom{\frac12} +
    \frac{3\, k}{2\pi U_2}\,\Li_4(q^{k}) +\frac{3}{4\pi^2 U_2^2}
    \,  \Li_5(q^{k})+c.c.\Big]\bigg\} \,,
\ee
where the torus is defined as a parallelogram (with edges identified) with sides of length $L_1$ and $L_2$ and angle $\theta$. Here $U := U_1+i U_2= (L_2/L_1) \, e^{i\theta}$ is its dimensionless complex structure and $\cV = L_1 L_2 \sin \theta$ is its volume, while $q := e^{2\pi i U}$, $\zeta(z)$ is the Riemann zeta function and the poly-logarithm functions are defined by
\begin{equation}
    \Li_\sigma(x) = \sum_{n=1}^\infty \frac{x^n}{n^\sigma} \,.
\end{equation}
It is the overall pre-factor $\cV^{-2}$ of this result that sets the scale of its size to be order $M_\KK^4$ because the complex structure $U$ involves only dimensionless quantities. This scaling with extra-dimensional size is also true for massless extra-dimensional fields in a variety of other geometries \cite{KKCase}.

The extra-dimensional theories in which such calculations are performed are {\it not} UV finite and the Kaluza-Klein calculation of $\rho_4$ generically diverges in the UV (though not at one loop in odd dimensions, as it turns out). For instance \pref{6to4masslessC} is obtained by taking the $m \to 0$ limit of the mode sum\footnote{Since \pref{6to4masslessC} also involves a double sum it might not seem to represent much progress over \pref{2torus}, but the point is that the sums in \pref{6to4masslessC} converge very quickly because they are organized as series in $|q| = e^{-2\pi U_2}$.}
\be \label{2torus}
    \rho_4 = \mu^{4-d}  \sum_{k,l\in\bZ} \int \frac{\exd^dp}{(2\pi)^d}
    \ln\left[ \frac{p^2 + M_{kl}^2 + m^2}{\mu^2} \right]  
    =  -\frac{\mu^{4}}{(2\pi)^{d}} \sum_{k,l \in\bZ} \int_0^\infty \frac{\exd t}{t^{1+d/2}} e^{-\pi\,t\, \big[\,(M_{kl}^2+m^2)/\mu^2\big]} \,,
\ee
where $M_{kl}$ denotes the Kaluza-Klein spectrum
\be\label{mass22}
    M^2_{kl}(\sigma_1,\sigma_2) =   \frac{(2 \pi)^2}{\cV\, U_2} \, \Bigl| l +\sigma_{2}-U
    (k+\sigma_{1})\Bigr|^2 \,.
\ee
with $0 \leq \sigma_1,\sigma_2 \le 1$ measuring the twisting\footnote{For instance $\sigma=\frac12$ if a field is taken to be antiperiodic rather than periodic around one of the toroidal cycles.} of the boundary conditions around the cycles of the torus (and it is the special case $\sigma_1 = \sigma_2 = m = 0$ that gives expression \pref{6to4masslessC}). 

Because this sum is so explicit one can explore its actual sensitivity to UV scales and see why the final result is often small. Although the sums converge in the final expression in \pref{2torus} -- and can be performed explicitly in terms of Jacobi theta functions -- the UV divergence shows up when integrating over the heat-kernel parameter $t$, which does not converge at the $t \to 0$ end. This is regularized above using dimensional regularization, with the complex quantity $d = 4 - \epsilon$ ultimately taken to 4. In \pref{2torus} $\mu$ is the usual arbitrary dim reg mass scale that ultimately drops out of all physical quantities.

The ultraviolet divergent part of \pref{2torus} can be identified very explicitly by tracking the pole as $\epsilon = 4-d \to 0$, and is given (for all $\sigma_1$ and $\sigma_2$ and $m$) by 
\begin{equation}\label{uvdivergent}
    \rho_{4\infty} = \frac{m^6 \, \cV}{192 \pi^3 \epsilon} \,.
\end{equation}

This expression has several noteworthy features. 
\begin{itemize}
\item It depends on the moduli, $L_{1,2}$ and $\theta$, of the toroidal geometry only through the volume $\cV$. This ensures it contributes to $\rho_4$ in the same way as would a 6D cosmological constant.  
\item It is proportional to $m^6$, also consistent with what would be expected in dimensional regularization for a divergent contribution to the 6D cosmological constant. 
\item It is $\sigma_i$-independent (and so independent of the boundary conditions in the extra dimensions). 
\end{itemize}
All of these features reflect the fact that this $1/\epsilon$ pole represents a {\it bona fide} 6D divergence, despite it emerging as $d \to 4$ and within the context of an apparently 4D calculational framework where the vacuum energy is computed mode-by-mode. Although the dimensional continuation arose by deforming to nonzero $\epsilon = 4-d$ in the $p$-integration, this deformation also plays a role in other manipulations (such as the interchange of summation and integration) so $\epsilon \neq 0$ indirectly regularizes these as well. Indeed, the resulting divergent part is identical to what is found starting from 6D and following the powers of $\epsilon' = (6-d)$ using general 6D short-distance heat-kernel expansions \cite{Hoover:2005uf, Burgess:2005cg}.

It is because the divergence is short-distance in 6D that \pref{uvdivergent} is so simple. The divergence comes from short-wavelength modes in {\it all} six dimensions, and so they only `see' the local properties like local curvatures and cannot be sensitive to global properties like boundary conditions. 

It happens that the absence of other powers of $m$ and $\cV$ is an artefact of the spacetime being flat, but if repeated for other curved geometries (such as compactification on spheres) there are more divergences that can depend on powers of $m$ and extra-dimensional size $L$ as $m^6 L^2$, $m^4 \log (mL)$, $m^2/L^2$ and $1/L^4$, precisely as would arise from a local contribution to the action involving (in six dimensions) up to three powers of curvature invariants; schematically:
\be \label{UVlocal}
   S_\UV = \int \exd^6x \; \sqrt{-g} \Bigl(c_0 m^6 + c_1 m^4 \cR + c_2 m^2 \cR^2 + c_3 \cR^3 \Bigr) \,,
\ee
with the extra-dimensional volume contributing $L^2$ and each curvature contributing $\cR \propto L^{-2}$. It is indeed because of this structure that the divergences can be renormalized into counterterms in the extra-dimensional theory.\footnote{Additional UV-sensitive contributions -- that in this case {\it can} depend on the boundary conditions -- occur for orbifolds \cite{Ghilencea:2005vm}, corresponding to new counterterms localized at the orbifold points, but otherwise the argument is the same.}  

Similar statements apply to the $m$-dependence of the UV-finite parts of the calculation. Using the full result \pref{2torus} for $\rho_4$ on a torus only $m^6 \cV$ arises, plus corrections that are exponentially small in the limit $m L \gg  1$. For the case with nonzero background curvature general heat-kernel expansions \cite{Hoover:2005uf, Burgess:2005cg} give a series in powers of $mL$, again consistent with a local curvature expansion involving terms of the schematic form  $\int \exd^2x \, m^6 (\cR/m^2)^n$.  

This exploration of UV sensitivity shows precisely when and why vacuum energies obtained by KK sums can (but need not) be UV insensitive. Calculations for tori are particularly simple because the absence of background curvature precludes any UV scales from appearing in a way that is not proportional to $\cV$, leading to a result that is always given by the tower spacing: $\rho_4 \sim M_\KK^4 \sim L^{-4}$. For more general geometries the same is true for {\it massless} fields in the extra dimensions, again for want of another scale to combine with $L$. 

But for massive extra-dimensional fields in curved backgrounds Casimir energies in general are more complicated and can be dominated by extra-dimensional UV scales rather than simply by the tower spacing, although non-negative powers of UV scales like $m$ only arise in a way that is consistent with the theory's local counter-terms within the extra dimensions. This UV dependence can be dangerous -- but need not be, as experience (for instance) with large extra dimensional models \cite{Arkani-Hamed:2000hpr, SLED} shows. Viability of extra-dimensional approaches to the cosmological constant problem includes providing a mechanism for why such UV sensitivity drops out: What solves the {\it higher-dimensional} cosmological constant problem? For SLED models the mechanism is the supersymmetry of the bulk (amplified by the accidental scaling symmetries generic to supergravities in 6 or more dimensions \cite{UVShadows}). For these models bulk supersymmetry plays two related roles: it enforces cancellations of UV effects amongst the contributions of massive bulk fields within a 6D supermultiplet \cite{Burgess:2005cg}, and it also forbids some local counterterms (like the extra-dimensional cosmological constant\footnote{Minimal supersymmetry in six or more dimensions can forbid higher dimensional cosmological constants \cite{Salam:1989ihk}, much as multiple supersymmetries can do in 4D.} itself.) 

\subsubsection*{Conjectures}

Ref.~\cite{Montero:2022prj} says that the swampland distance conjecture ensures that 4D vacuum energies in the presence of towers of states (with spacing $m$) require the scaling
\be \label{SwampThing}
  \rho_4 \sim m^p \qquad \hbox{for some positive power $p$}. 
\ee
So far as the cosmological constant problem goes, the essence of \pref{SwampThing} is that UV physics will have to involve a tower whose spacing is set by the eV scales relevant to dark energy. This kind of spacing is indeed present in extra-dimensional approaches to the cosmological constant problem, for which much effort has been invested in computing vacuum energies. This makes them useful benchmarks against which to compare the newer conjectures, eventually allowing an assessment about  whether more sweeping assumptions are necessary.

At its weakest \pref{SwampThing} is just the statement that $\rho_4$ should vanish in the limit $m \to 0$, as would be expected in a higher-dimensional realization if higher-dimensional flat space must be a solution to the extra-dimensional theory. (This is often true, but has a robust EFT understanding in terms of the accidental scaling symmetries of string vacua \cite{UVShadows}.) Although \pref{SwampThing} is satisfied by most of the terms in \pref{UVlocal}, it is not true for the first few. Its failure is consistent (for example) with an extra-dimensional cosmological constant precluding maximally symmetric flat space being a solution to the field equations. 

A stronger interpretation of \pref{SwampThing} instead is that the numerical size of $\rho_4$ is given purely by a power of the KK scale (as opposed to higher UV scales like brane tensions or the string scale). This also has extra-dimensional counter-examples within concrete extra-dimensional models, such as when $\rho_4$ is controlled by the tensions of space-filling 3-branes in an extra-dimensional geometry. Indeed this is one of the reasons the non-supersymmetric 6D story \cite{Carroll:2003db} ultimately fails, since perturbing the 3-brane tensions by $\delta T$ in the initially 4D-flat scenario ends up introducing a 4D cosmological constant of size $\delta T$ \cite{TensionIsRho}.

At face value one of two conclusions seems inevitable: either the conjectures behind \pref{SwampThing} are wrong, or the various KK counterexamples should be cast into the swamp. In order to judge the utility of the conjectures it would be very useful to have a concrete extra-dimensional model that is both anointed in advance as not being in the swampland, and sufficiently concrete that its implications could be explicitly explored (as required in order to test UV robustness in a meaningful way).

But even if \pref{SwampThing} proves false, towers still contain some magic since Casimir energies for massless extra-dimensional fields really are set by KK scales, and it remains true that all but the lowest few members of these towers are much heavier than this. Indeed, the observation that Casimir energies from KK towers of massless fields are set by the KK scale plays an important role in extra-dimensional models, because it is usually hoped that it is the Casimir energies of these fields that ultimately survive to provide a nonzero result for $\rho_4$ after whatever suppresses contributions at UV scales has done its work. This is ultimately the reason why the extra-dimensional size in these models is chosen with $M_\KK$ in the eV range.\footnote{A precise statement for the exact size required of the KK scale is usually difficult because it requires a reliable calculation of the 4D vacuum energy to an accuracy that includes the subdominant terms that survive once the naively dangerous leading UV physics has successfully been removed.}

\subsection{Breakdown of EFT reasoning}

How surprised should we be to find that towers can in some circumstances have reduced vacuum energies (in the cases like massless extra-dimensional fields or massive fields on tori)? Does the disagreement between the tower result and a naively truncated level-by-level calculation represent a significant breakdown of EFT reasoning? We now show -- following arguments made in \cite{EFTBook} -- why it does not. The main point is that there is no reliable EFT estimate of the vacuum energy that includes only a finite number of nonzero levels in a tower, so EFTs are mute about what the result should be. They are mute because for towers EFT reasoning breaks down in a very mundane way: the underlying hierarchy of scales that EFTs assume does not exist. There is no inconsistency with EFT reasoning provided one is clear about its domain of validity. 

What makes EFTs useful is that they tell you in advance where they must fail: they assume the existence of a hierarchy of scales -- degrees of freedom with energy $E_{\rm high}$ are integrated out in order to better understand other degrees of freedom with energy $E_{\rm low} \ll E_{\rm high}$. Commonly assumed properties rely on it being a good approximation to work order-by-order in $E_{\rm low}/E_{\rm high}$, including the very locality of the EFT itself. Locality depends on this in detail because it is only after heavy propagators are expanded to fixed order in $p^2/M^2$,  
\be
   \frac{1}{p^2 + M^2} = \frac{1}{M^2} \left[ 1 - \frac{p^2}{M^2} + \cdots \right] \,,
\ee
that they become polynomials in momentum and so become capturable by local operators built from fields and their derivatives. 

Returning now to towers, suppose the mass of the {\it lightest} state of the tower to be integrated out is $M$ and the mass of the {\it heaviest} tower state in the low-energy sector is $m$. EFT methods explicitly assume the validity of an expansion in powers of $m/M$, and such an expansion can work well if the only level of the tower kept in the low-energy theory is the massless one. In this case the lightest massive state to be integrated out has a mass set by the spacing $M$ within the tower, while states in the low-energy EFT have masses $\delta m$ set by the size of any `fine-structure' splitting that generates splittings among the would-be massless level of the tower.\footnote{For example, if the tower consists of KK states the would-be massless states could include any moduli and the `fine structure' would come from any modulus-stabilization effects.} Corrections to leading order reasoning in this case are controlled by the ratio $\delta m/M$, which can be small. This is indeed how extra-dimensional field theories emerge as the low-energy description for string vacua and also how 4D EFTs emerge as the low-energy limit of higher-dimensional theories. 
 
For a tower of states what is {\it never} a good approximation is to include a level of the tower with nonzero mass into the low-energy theory and integrate out the higher tower levels. To see why, suppose the zeroth order level spacing is given by $M_\ssN^2 = N \mu^2$ (as happens for string levels), and suppose we keep the $N = N_c > 0$ state in the low-energy theory but integrate out all states with $N \geq N_c+1$. In this case 
 \be
     \frac{m}{M} = \sqrt{\frac{N_c}{N_c+1}} 
\ee
which for $N_c = 1,2,\cdots$ is at its {\it smallest} when $N_c = 1$, at which point $m/M = 2^{-1/2} \simeq 0.707$. Any attempt to include a fixed number of nonzero-mass tower states within the low-energy EFT comes with an explicit warning: corrections are never under calculational control. This is in particular true for predictions about the vacuum energy for such a theory. 

Indeed, this observation is part of the reason why infinite towers of states are so interesting: they provide an explicit regime where a naive low-energy EFT description that keeps only a finite part of the tower is explicitly unavailable. Since our standard reasoning does not apply, surprises might conceivably lurk in the UV limit of such theories (such as did the UV finiteness of string theory itself).

That does not mean that EFT methods themselves are breaking down in a new or surprising way; EFTs never claimed to apply in situations without a hierarchy of scales. In the case of a Kaluza-Klein tower the correct result for $\rho_\ssD$ is obtained from within an EFT, but the EFT in question is the higher-dimensional field theory that describes the entire tower. So far as we know calculations involving towers of string states must be done within the full string theory.

\subsubsection*{Broader implications for `species' scales}

The previous section argues that if we find a tower of states when working our way up in energy then as soon as we hit the first nontrivial rung we have necessarily hit the UV cutoff of any EFT description that does not include the effects of the entire tower. 

At some level this is not a surprise: a 4D EFT necessarily breaks down at or below the KK scale and a higher-dimensional EFT always fails at or below the string scale. These are both special cases of a well-known fact in semiclassical gravity: in all of the examples understood in detail (which in practice limits us to the semiclassical regime) the UV cutoff in a gravity theory is much smaller than the Planck scale. In the 4D world $M_\KK$ is always smaller than the extra-dimensional Planck scale within the weakly curved regime accessible to semiclassical methods. In a higher-dimensional world with a stringy provenance the UV completion scale is (at most) the string scale, $M_s$, and this is much smaller than the higher-dimensional Planck scale in the regime of weak string coupling. 

Having towers be beyond the pale of EFT methods might have implications for some aspects of the swampland program, which aims to go beyond the explicit KK and string examples \footnote{The motivation for swampland conjectures is very ambitious: use experience with string theory to identify general properties of quantum gravity that could apply to any other candidate UV completions of gravity. Our aims are more limited: for want of alternatives we work with the known EFT framework of string and extra-dimensional models. We conjecture that this is not a restriction (which agrees with the emergence conjecture when applied to theories satisfying the distance conjecture). }. One thread within this tapestry identifies a `species' scale, $\Lambda_s$, that quantifies how much lower the UV cutoff might be than $M_p$ in a way less tied to the above concrete examples. $\Lambda_s$ seems to have different usages in different parts of the literature, but is sometimes suggestively chosen to lie part way up a tower \cite{Palti:2019pca, vanBeest:2021lhn, Agmon:2022thq}. The above arguments might make it worth revisiting conclusions that rely on this in an important way.\footnote{Or not. One interpretation for the species scale is that it is the lowest scale for which no possible low-energy EFT exists. Within string theory this is typically the string scale, which is usually much larger than the splittings amongst any lower-energy KK towers. This is consistent to the extent that the EFT below this scale is extra-dimensional and so contains the complete tower. }

\begin{figure}[t]
\begin{center}
\includegraphics[width=120mm,height=75mm]{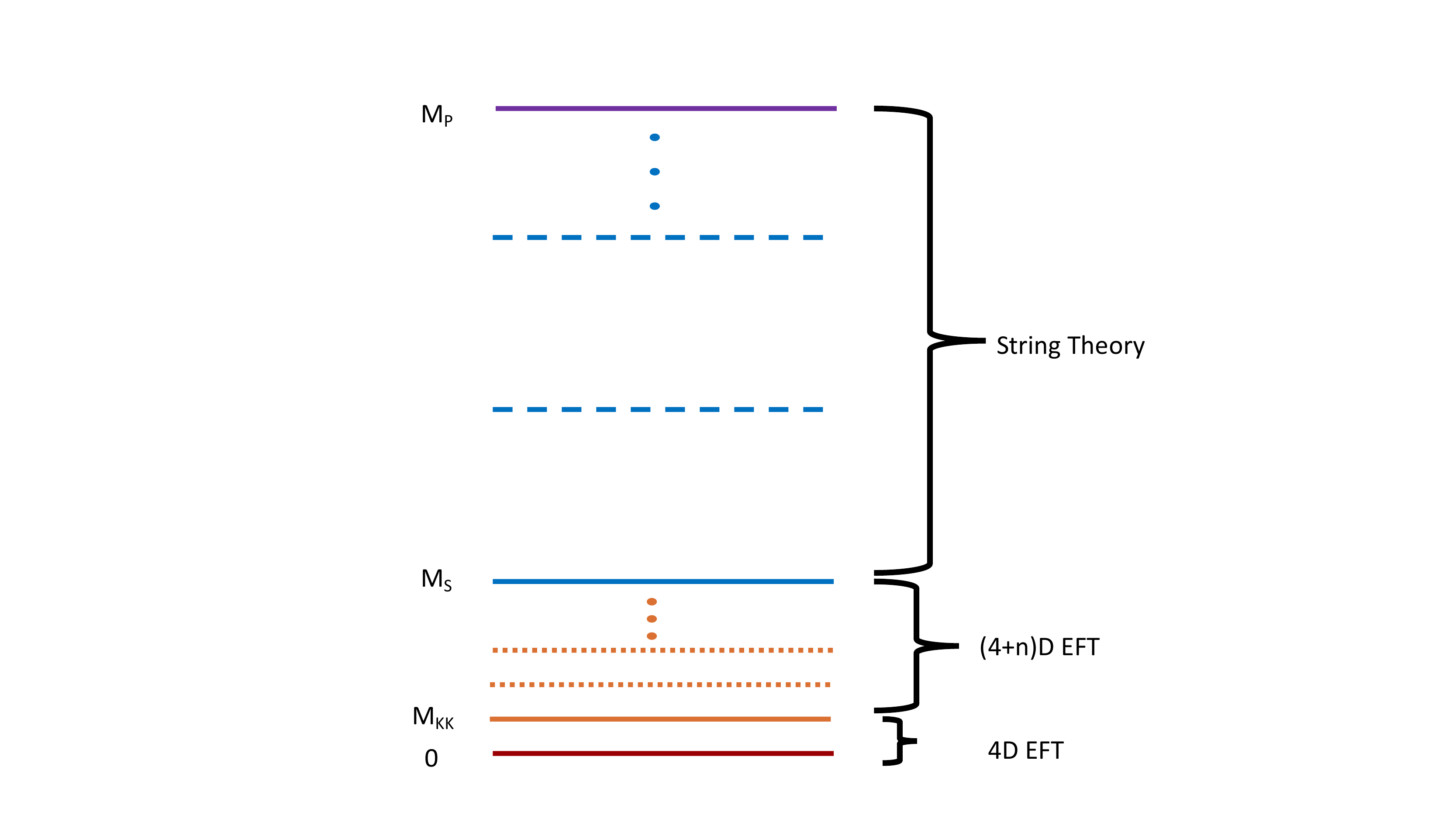} 
\caption{ An illustrative cartoon regarding the different cut-offs (or species scales) for Kaluza-Klein and string towers. Notice that the species scale has to be the lowest value of the corresponding tower which is the Kaluza-Klein and string scales respectively. Proposing a species scale only part way up the tower (as for instance any line representing a scale  between $M_{KK}$ and $M_S$) would give rise to an EFT which is not under control.} 
\label{Fig:Tower} 
\end{center}
\end{figure}

\section{Dark dimensions}

Does the suppression (in some circumstances) of tower vacuum energies provide a mechanism for solving the cosmological constant problem? No concrete proposals have yet been able to do so, though this mechanism is known to play at least a supporting role in those that rely on extra dimensions. 

In essence, the cosmological constant problem asks why everyday particles ({\it e.g.}~the electron) seem to contribute much more to the vacuum energy than is seen to be gravitating in cosmology. The hard part when trying to solve it is to suppress the vacuum energy while not ruining other things we know about low-energy physics. The electron is a useful benchmark when assessing any proposal: Where is it in the theory? Do its properties agree with the many things we know about electrons? Why is its vacuum energy suppressed by more than 30 orders of magnitude relative to what its mass suggests? 

At first sight the tower proposal seems to be a non-starter: to apply it to Standard Model particles seems to require these should live amongst tower states (so they can benefit from the suppression). But the tower shouldn't be split by more than eV energies so that the tower spacing can be of order the observed Dark Energy density. It is hard to see how experimenters could have missed a tower of electrons spaced by eV scales. But if Standard Model fields are not in the tower, then why is it central to their not contributing to the vacuum energy?

To avoid problematic proliferation of KK Standard Model states all models with eV sized KK scales (including \cite{Montero:2022prj}) postulate that the Standard Model is localized on a space-filling 3-brane situated within the extra dimensions. This is ultimately why extra-dimensional approaches are attractive: it allows non-gravitational electron physics to remain in 4D (and hopefully remain unchanged), while specifically changing only how it -- and its vacuum energy -- gravitates. But it makes calculating the vacuum energy much more subtle because the dangerous Standard Model vacuum energy is then a contribution to its local brane tension, so computing its implications for the 4D cosmological constant involves computing the back-reaction of brane physics on the bulk. Until this is done one really doesn't know what the predicted vacuum energy is. 

\subsubsection*{Back reaction}

A fair bit of work studies back-reaction for space-filling 3-branes within 6D theories, and it starts out with good news: codimension-two branes often curve the transverse dimensions rather than the on-brane dimensions their internal cosmologists would measure (and this observation drove the study of the simplest solutions \cite{Carroll:2003db, SLED}). This turns out to remain true for a broad class of classical solutions \cite{ClassicalSLED}, but not all (including known de Sitter solutions \cite{TensionIsRho, SLEDdS}). 

Part of what makes this hard is that it also requires a detailed picture of the modulus stabilization that fixes the size (and shape) of the extra dimensions. The physics of back reaction shapes both the size of the extra dimensions and the curvature of the 4D world (and so, effectively, the 4D vacuum energy). Explicit examples exist (for instance non-supersymmetric rugby balls \cite{TensionIsRho}) for which the 4D curvature is controlled by brane tensions rather than KK scales, and the underlying scale-invariance of extra-dimensional supergravity makes this direct connection between brane properties and the 4D cosmological constant very robust \cite{CCisBoundary}. 

So the hard part for a successful model is to understand why the 4D vacuum energy should be more insensitive to brane properties. It turns out that quantization of extra-dimensional fluxes plays a crucial role in making the stabilized extra dimensions fairly rigid, and this is largely responsible for passing on changes in brane tensions to the 4D geometry. A big step forward was seeing how higher-dimensional flux quantization gets captured in the low-energy 4D EFT, and it turns out this is done through the presence of 4-form flux fields \cite{SLED4DEFT} (suggesting these are likely also important ingredients\footnote{It is perhaps not surprising from this perspective that 4-forms also turn out to play a role in other approaches to the cosmological constant problem \cite{4FormCC}.} in the ultimate cosmological constant story). 

In SLED models a combination of extra-dimensional supersymmetry and scale-invariance helps decouple brane scales from $\rho_4$ \cite{SLEDSUSY}, but in the end no explicit examples seem to reduce the result by nearly enough (for the most explicit calculations see \cite{SLEDNotEnough, SLEDNotEnough0}). Our current thinking as to why 6D models have not yet completely succeeded \cite{Brax:2022vlf} is that the flux-based modulus-stabilization mechanism they use \cite{SalamSezgin} turns out to break the underlying string-based accidental scale invariances described in \cite{UVShadows} that favour flat solutions. This has led us to develop alternative stabilization mechanisms that do not break these symmetries \cite{Burgess:2022nbx}, and to explore the 4D implications to which such stabilizations could lead \cite{YogaDE}. While these approaches do make progress reducing the vacuum energy, the biggest question mark in this line of research seems to be the ubiquitous appearance of very light dilatons in the low energy theory (organically associated with the accidental scale invariance), whose couplings to matter seem too large to have been missed. Work is ongoing \cite{Brax:2022vlf, Burgess:2021qti} to see how fatal a problem this really is.  

\subsubsection*{5D vs 6D}

It is early days for making detailed comparisons contrasting how 5D and 6D models handle the back-reaction issue, but we round out this section by highlighting some positives and negatives of the two approaches. Contrasting the newer 5D approach with the older 6D model in this way is useful since the 6D model provides a benchmark for what is known to be possible using standard EFT methods (and their limitations), and so provides a yardstick against which alternatives can be compared.\footnote{We explore here only unwarped models with large dimensions.} 

A big difference between 5D and 6D large-dimension models is the relation they predict between the KK scale, $M_\KK$, the 4D Planck scale, $M_p$, and the higher-dimensional gravity scale $M_5$ or $M_6$. Ignoring order unity factors these are related by the well-known relations \cite{ADD}
\be
    M_5 \sim \Bigl( M_p^2 M_\KK \Bigr)^{1/3} \qquad \hbox{and} \qquad M_6 \sim \Bigl( M_p M_\KK \Bigr)^{1/2} \,,
\ee
which for $M_\KK \sim 1$ eV and $M_p \sim 10^{18}$ GeV imply $M_5 \sim 10^9$ GeV and $M_6 \sim 3 \times 10^4$ GeV. Neither of these is currently ruled out as an extra-dimensional model \cite{ADDBounds}, though in the 6D case order-unity factors matter because the most robust model-independent bounds\footnote{One sometimes sees much more aggressive bounds quoted for the 6D case -- {\it e.g.}~\cite{RaffeltStrong} -- but these all rely on KK modes decaying into photons and so are more model-dependent. They can be evaded, for instance, by coupling KK modes more efficiently to a non-Standard Model sector (sometimes colourfully called the `toilet' brane) to depress their branching ratio into photons \cite{MSLED}.} (energy loss from hot astrophysical bodies) constrain $M_6 \gsim 30$ GeV.

We finally briefly comment on what kinds of ingredients both approaches might require from the point of view of a UV completion into string theory. Both assume a KK scale determined by the value of the cosmological constant, and so both require a quasi de Sitter compactification with a large volume of the extra dimensions as in \cite{LVS, RG}. Because all dimensions cannot be this large both require this compactification to be stabilized in a anisotropic way (as is done in \cite{BCQ} for the 6D case). 

In both cases the Standard Model must be localised on nonsupersymmetric branes, such as a non-supersymmetric D3 brane or D7's wrapping a small cycle as in F-theory. Both are consistent with swampland conjectures. The absence of an explicit string theory realisation for either case is an interesting challenge (see however \cite{Grimm:2013fua}). The fundamental (string ) scale is the TeV scale for 6D and $10^9-10^{10}$ GeV for the 5D case.\footnote{See \cite{Luis} for another proposed set of scales based also on swampland arguments.}

As mentioned above, both scenarios seem to be broadly viable from a phenomenological point of view. In both cases the size of the large dimensions is by construction put above the current experimental bound \cite{Kapner:2006si, Adelberger:2003zx}. The other model-independent constraints restrict the fundamental scale and for the 6D case model building is required to exclude the photon as the dominant decay channel for the extra-dimensional KK modes. No such evasion seems required for the 5D scenario, though the phenomenological study of this scenario is less well developed. 

The main difference is the detailed studies of back reaction effects from non-supersymmetric high scale branes to the bulk that are available in 6D. To our knowledge these issues have not yet been addressed in 5D. The proof of the pudding is in the explicit calculation of the cosmological constant, which can be done in the 6D (with, so far, insufficient suppression) and which has not been done at all in 5D. This remains the main obstacle to both approaches so far, and much work remains to see which performs the best once the two approaches are eventually put on the same footing.

\section{Conclusions}

Exploring the large extra dimensions scenario in order to address the dark energy problem is a promising avenue, either motivated by general properties of EFTs or by swampland conjectures. 6D and 5D proposals share interesting properties, and furnish well-defined alternatives to the anthropic proposal of Weinberg-Bousso-Polchinski \cite{Weinberg:1987dv, Bousso:2000xa} (see \cite{PhysRep} for a recent review) with the potential bonus of leading to concrete low-energy predictions that could be tested experimentally, both at colliders and different tests of gravity. Even though, at present, neither reaches the ambitious target to explain the small size of the measured dark energy -- including all potential quantum effects -- from a UV complete theory, it is in principle possible and testable.

\begin{table}[ht]
\centering 
\begin{tabular}{|c|c|c|} 
\hline
\cellcolor[gray]{0.9}{\bf Yardstick} & \cellcolor[gray]{0.9} \bf 5D & \cellcolor[gray]{0.9}\bf{6D}  \\ 
\hline 
\hline
Kaluza-Klein  Scale & $\sim 1$ eV & $\sim 1$ eV  \\
\hline
Implied Fundamental  Scale & $10^9-10^{10}$ GeV & $10^4$ GeV  \\ 
\hline
Consistent with Swampland Conjectures & Yes & Yes \\
\hline
Localised Standard Model & Yes  & Yes \\
\hline
Bulk SUSY Breaking  Scale & ? & $\sim 1$ eV\\
\hline
Brane SUSY Breaking  Scale & $ 10^3-10^9$ GeV & $ 10^4$ GeV \\
\hline
Modulus Stabilisation Mechanism? & Not yet & Yes \\
\hline
Explicit String Realisation & Not yet & Not yet \\
\hline
Viable Phenomenology & Yes & Yes* \\  
\hline 
Higher-Dimensional EFT & Possibly & Yes \\
\hline
Four-Dimensional EFT & Possibly & Yes \\
\hline
Allowed Bulk Cosmological Constant? & Possibly & No \\
\hline
Computable Vacuum Energy & Not yet & Yes  \\
\hline
Ability to Reproduce Observed Energy Density  & Not yet & Not yet \\
\hline 
\end{tabular}
\caption{Comparison of the level of development between the 5D and 6D dark dimension scenaria. The asterisk indicates the existence of model-dependent bounds contingent on KK modes decaying significantly into photons, whose evasion is possible but requires model building.}
\label{Tab5D6D}
\end{table}

Table \ref{Tab5D6D} summarizes at a glance similarities and differences of the 5D and 6D proposals. 

\medskip

In this note we convey several ideas:

\begin{itemize}
\item We show why for towers of states a scaling relationship between 4D vacuum energy and the tower spacing $m$ of the form $\rho_4 = m^{p}$ considered in \cite{Montero:2022prj} can (but need not) arise naturally within controlled EFTs describing KK towers. In generic compactifications with nonvanishing curvature the coefficients in these scaling relations can (but need not) depend on higher-dimensional UV scales as well as on $m$. The EFT analysis is instructive because it also tells you why and when these kinds of relations hold.
\item If a tower of states is hit as one climbs in energy then the first nontrivial level is as high as the low-energy EFT can possibly be applied. For instance, a lower-dimensional analysis of 4D states in the middle of a KK tower is never under EFT control.
\item Proposals using large dimensions to explain dark energy never allow the Standard Model to live in the extra dimensions and so don't live in a KK tower in the large dimensions (being instead localized on a brane, say). The special properties of tower vacuum energies are not then directly relevant to their contribution to the vacuum energy (the cosmological constant problem). Instead, these models make the dark energy problem into an exercise for which back reaction becomes the important question. One must re-ask the cosmological constant problem question in the higher dimensions: why do UV scales like brane tensions (which cannot be BPS because of the lack of supersymmetry) not source the vacuum energy. (This is the hard part of the problem, of course, and seems not yet addressed in detail in 5D.)
\item The new 5D and older 6D proposals seem equally consistent from the swampland point of view, inasmuch as they both seem to satisfy the relevant conjectures (though we do not claim to be up-to-date with the modern canon). But are broadly phenomenologically viable against model-independent experimental tests. 5D models are less tightly pressed because it involves a higher extra-dimensional gravity scale. 6D models require model building to ensure that the leading decay mode for KK modes is not into photons (which can be done by providing them with more efficient decays into invisible final states). %
\item 6D models mostly benefit from having been studied in more detail so the back-reaction problem is better understood there, at least for some stabilization mechanisms. This experience might be useful for exploration of the 5D models. 
\end{itemize}

Most importantly, no 5D or 6D model has a completely satisfactory calculation of the vacuum energy or resolution of the cosmological constant problem, and so all are best regarded as works in progress. 

\section*{Acknowledgements}
We thank Sebasti\'an C\'espedes, Shanta de Alwis, Chris Hughes,  Francesco Marino, Miguel Montero, Francesco Muia,  Veronica Pasquarella, Mario Ramos, Cumrun Vafa, Irene Valenzuela and Gonzalo Villa for helpful conversations. FQ thanks Perimeter Institute for hospitality during the development of this work.
CB's research was partially supported by funds from the Natural Sciences and Engineering Research Council (NSERC) of Canada.  
The work of FQ has been partially supported by STFC consolidated grants ST/P000681/1, ST/T000694/1. Research at the Perimeter Institute is supported in part by the Government of Canada through NSERC and by the Province of Ontario through MRI.

\end{document}